\documentclass[reprint,aps,prl]{revtex4-1}
  
\usepackage{graphicx}
\usepackage{amsmath,amssymb,amsfonts,mathrsfs}
\usepackage{hyperref}
\usepackage{siunitx}
\usepackage{oplotsymbl}
\definecolor{blue}{HTML}{0072BE}
\definecolor{green}{HTML}{77AD30}
\definecolor{pink}{HTML}{FF0080}
\definecolor{yellow}{HTML}{EEB220}
\definecolor{red}{HTML}{DA5319}
\definecolor{marron}{HTML}{A3142F}

\begin{document}

\title{Pressure-induced densification of vitreous silica: insight from elastic properties}

\author{Coralie Weigel}
\author{Marouane Mebarki}
\author{S\'ebastien Cl\'ement}
\author{Ren\'e Vacher}
\author{Marie Foret}
\author{Benoit Ruffl\'e}\email[Corresponding author:$\;$]{benoit.ruffle@umontpellier.fr}

\affiliation{Laboratoire Charles Coulomb (L2C), University of Montpellier, CNRS, Montpellier, France.}

\date{\today}  

\begin{abstract}
\textit{In situ} high-pressure Brillouin light scattering experiments along loading-unloading paths are used to investigate the compressibility of vitreous silica. Below \SI{9}{GPa}, the equation of state obtained from the sound velocities corrected for dispersion agrees with volume measurements. Conversely, huge anelastic effects are observed in the range \SIrange{10}{60}{GPa}, unveiling the reversible transformation from the fourfold-coordinated structure to the sixfold one. The associated density changes correlate with the average Si coordination number. Decompression curves from above \SI{20}{GPa} reveal abrupt backward coordination changes around \SIrange{10}{15}{GPa} and significant hysteresis. Contrary to common wisdom, the residual densification of the recovered silica samples can be figured out from changes in elastic properties along pressure cycles, ruling out a plastic description of the latter process.  
\end{abstract}
\maketitle
          
Changes in the structure of network-forming glasses occur in response to applied pressure $P$. Moreover, different amorphous states with distinct short- and/or intermediate-range orders as well as contrasting physical properties can be produced following separate thermo-mechanical paths. Even if these phenomena, termed {\em polyamorphism}, have been extensively studied during the last decades, their complete understanding still remain an ongoing challenge. It relates to the difficulty in carrying out conclusive experiments at high $P$. Beyond physics of the amorphous state, a comprehensive knowledge of polyamorphism has significance for the development of new functional glassy materials, particularly if novel high-pressure forms are recoverable to ambient conditions~\cite{MCMI2002, BRAZ2007}. 

Vitreous silica ($v$-SiO$_2$) is the archetypal tetrahedral network-forming glass whose behavior under pressure is of long-standing interest due to its primary importance as the analog material of silicates in geophysics~\cite{SANL2013, MILL2015}. At room temperature, the hydrostatic compression is reversible up to $\sim$\SI{10}{GPa}~\cite{MEAD1987,TSIO1998}. When compressed above this limit, the recovered glasses after complete unloading exhibit residual densification. The latter saturates at about 20\% for maximum pressures around \SIrange{20}{25}{GPa}~\cite{POLI1993}. \textit{In situ} experiments above \SI{25}{GPa} show that the density of the squeezed silica glass further increases gradually and becomes comparable to that of stishovite, the sixfold coordinated crystalline polymorph, above \SI{50}{GPa}~\cite{SATO2008}. The basic mechanisms of the SiO$_2$ network collapse upon compression have been early identified as i) the reduction of the Si$-$O$-$Si bond angle between SiO$_4$ tetrahedra below \SI{10}{GPa}, ii) the change in the ring size distribution, and iii) the progressive increase in the Si coordination number from four to six, the latter being not quenchable at ambient pressure~\cite{HEML1986, WILL1988, SUSM1991, MEAD1992, ZHA1994, SATO2008, BRAZ2009b, BENM2010, SATO2010, ZEID2014b, TREA2017, PRES2017}. 

Beyond structural studies, investigations of the thermodynamical and relaxational properties have proven to be enlightening. For instance, sound velocity measurements have revealed the anomalous negative pressure derivatives of elastic moduli at low $P$~\cite{KOND1981}. They have also provided an equation of state (EoS) $-$ density \textit{vs} pressure $-$ for $v$-SiO$_2$ below $\sim$\SI{10}{GPa}~\cite{SCHR1990, ZHA1994, WEIG2012, COAS2014}. The stiffening of the elastic moduli, associated with residual densification of the recovered glasses at ambient conditions, has also been evidenced many times~\cite{GRIM1984, POLI1990, ROUX2010}. Accurate volumetric experiments at pressures below \SI{9}{GPa} have revealed several features of polyamorphism in some other glasses among which logarithmic kinetics and significant inelastic effects are the more salient ones~\cite{TSIO1998, BRAZ2008, BRAZ2016}. However, most of the studies reported so far involve compression only. Although it is admitted that the high-$P$ octahedral structure of $v$-SiO$_2$ reverts back to the tetrahedral one at ambient conditions, little is known about this back transformation and its relation with other properties like density or elastic properties. The same is true for the residual densification process. In this Letter, we re-investigate the variations of sound velocities in $v$-SiO$_2$ using pressure cycles and \textit{in situ} Brillouin light scattering experiments of unprecedented quality. Combined with density data we provide a series of quantitative results related to the transformations along the loading-unloading paths.

High pressures were generated using Chervin-type diamond anvil cells (DAC) with cullets of \SI{400}{\um} diameter. Samples of about \SIrange[range-phrase = $\times$]{50}{50}{\um^2} were made from a \SI{15}{\um} thick polished plate of Tetrasil SE fused silica ([OH]~$\simeq$~100~ppm). They were each loaded in a chamber of \SI{150}{\um} diameter drilled in rhenium gasket together with ruby-spheres to measure the pressure~\cite{CHER2001}. The accuracy on pressure measurements was \SI{0.1}{GPa} and the pressure-transmitting medium was argon fluid to ensure an hydrostatic stress up to the highest $P$. \textit{In situ} high pressure Brillouin light scattering (BLS) experiments were performed using a standard triple-pass tandem interferometer~\cite{LIND1981} and a single line diode-pumped solid-state laser operating at $\lambda_0$ = \SI{532}{nm}. All the measurements were obtained at room temperature in the symmetric platelet geometry with a scattering angle of \ang{50}~\cite{WEIG2012}. The collection aperture was limited using a curved slit matching the spurious geometrical broadening of the Brillouin lines to the resolution of the spectrometer~\cite{VACH2006}. From the Brillouin frequency shifts, the velocity of the high frequency (\SIrange{5}{20}{GHz}) longitudinal and transverse acoustic waves, $v_\textsc{la}$ and $v_\textsc{ta}$ respectively, are obtained. These are related to the bulk modulus $K$ and to the shear modulus $G$,
\begin{equation}
v_\textsc{la}=\sqrt{\frac{K+\frac{4}{3}G}{\rho}} \;,\; v_\textsc{ta}=\sqrt{\frac{G}{\rho}},
\label{eq:v}
\end{equation}
where $\rho$ is the density. 

\begin{figure}
\centering
\includegraphics[width=8.6cm]{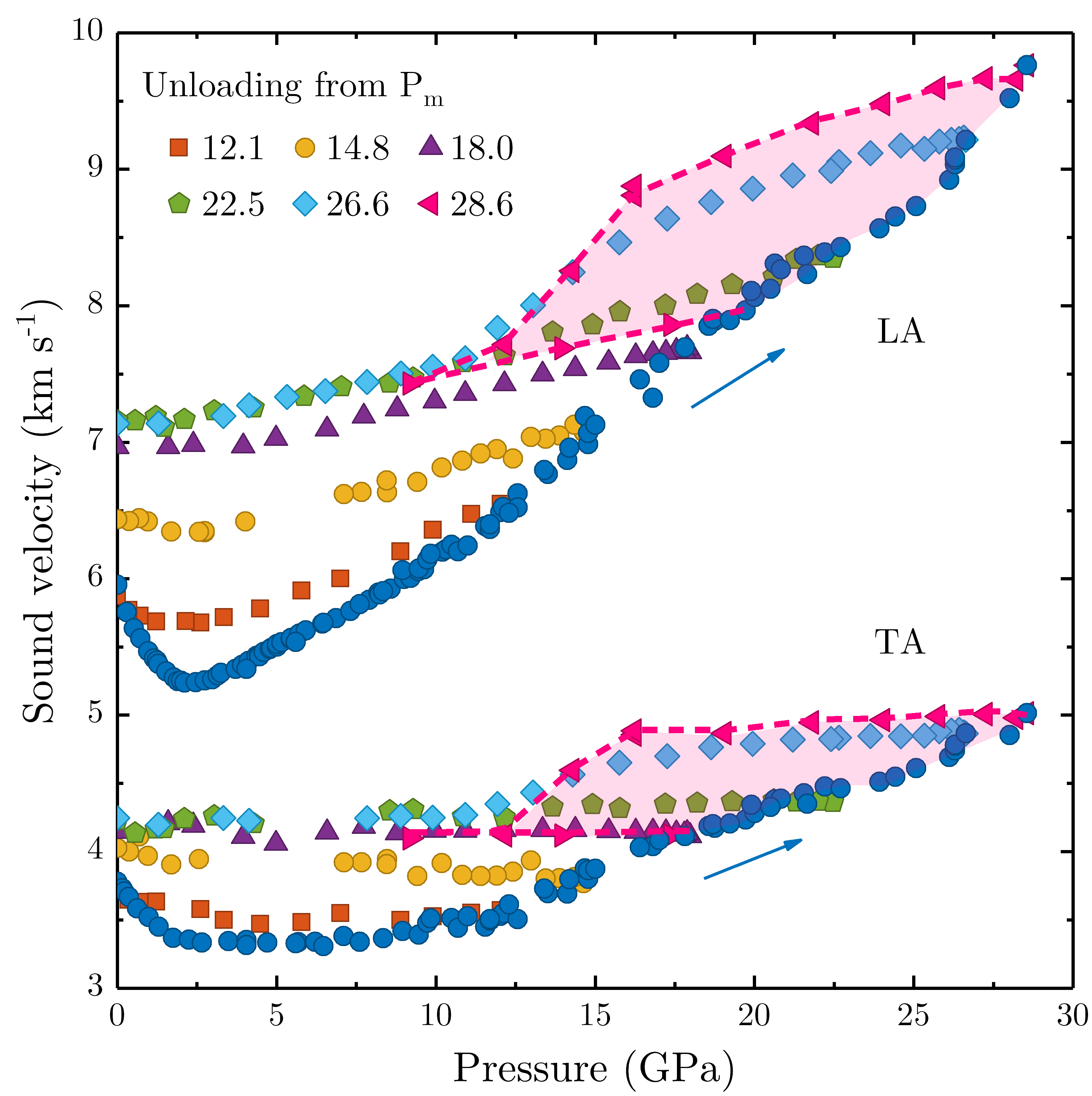}
\caption{Longitudinal (LA) and transverse (TA) sound velocities in $v$-SiO$_2$ upon compression (\circletfill[blue]) and during decompression from various maximum pressures $P_\text{m}$. The pink dashed line and the dashed area are guides for the eye.}
\label{fig:vLATA}
\end{figure}

Compression-decompression cycles reaching various maximum pressures $P_\text{m}$ extending up to nearly \SI{30}{GPa} have been carried out. As expected, the results found upon compression for both LA and TA velocities, shown as blue circles in Fig.~\ref{fig:vLATA}, superimpose on a single curve for all cycles. For $P_\text{m}\alt\SI{10}{GPa}$, the values of the sound velocities upon decompression also match the ones on compression within the accuracy of the measurements (See S-I and Fig.~S1 in Supplemental Material). The other symbols in Fig.~\ref{fig:vLATA} show the results acquired for decompression paths from higher $P_\text{m}$ ranging from \SIrange[range-phrase = ~to~]{12.1}{28.6}{GPa}. Upon decompression, the velocities are larger than those obtained in the compression part of the cycles, indicating strong modifications in the elastic properties, but also revealing unsuspected hysteresis phenomena at high $P$ (pink shaded area in Fig.~\ref{fig:vLATA}). These two points will be discussed later on.

\begin{figure}
  \centering
  \includegraphics[width=8.6cm]{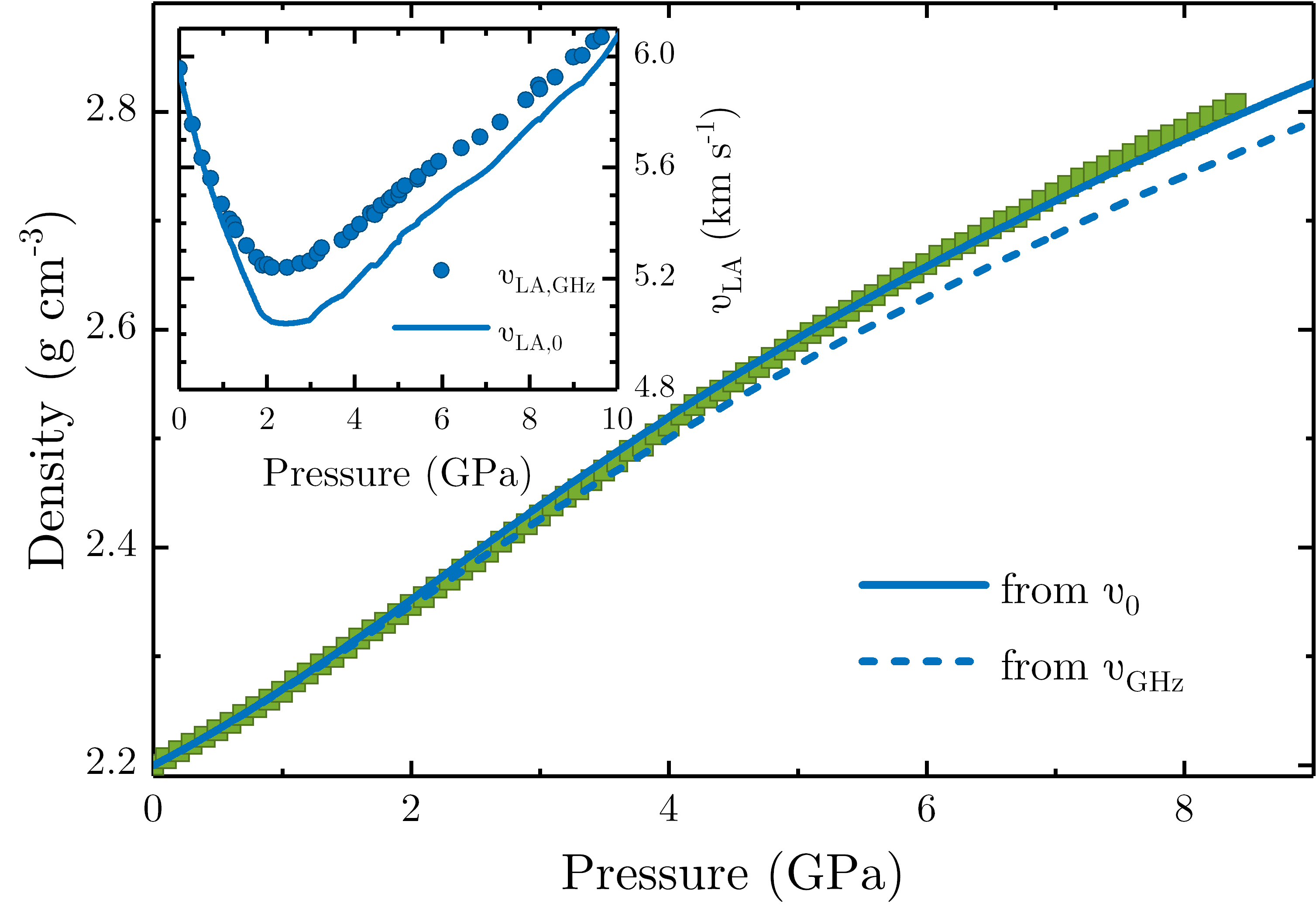} 
  \caption{Density of $v$-SiO$_2$ from volumetric experiments (\squadfill[green]~\cite{TSIO1998}), compared to values computed from Eq.~\ref{eq:rhoPvk} using the hypersonic velocities, as measured ($v_{\textsc{gh}\text{z}}$ dashed line) or corrected from dispersion effects ($v_0$ solid line). Inset: LA velocities $v_{\textsc{la,gh}\text{z}}(P)$ (\circletfill[blue]) and associated relaxed velocities $v_\textsc{la,0}(P)$.}
  \label{fig:Rho}
\end{figure}

We will focus our analysis on the quantity $\rho\chi$ where $\chi=K^\text{-1}$ is a compressibility. This quantity is actually of prime interest since it gives the infinitesimal density variation $\delta\rho$ associated with a pressure variation $\delta P$. Volumetric measurements give a direct access to the static compressibility $\chi_0 = -\frac{1}{V}\frac{\partial V}{\partial P} = \frac{1}{\rho}\frac{\partial \rho}{\partial P}$, and to the density $\rho$ using the mass conservation, thus leading to $\delta\rho = \left(\rho \chi_0 \right) \delta P$. The cumulative sum of the $\delta \rho$ is the density increase $\Delta \rho$ as a function of $P$. On the other hand, sound velocities are directly related to the $\rho\chi_\textsc{e}$ quantity, $\rho\chi_\textsc{e} = v^\text{-2}_\textsc{k}$, where $v^2_\textsc{k} = v^2_\textsc{la} - \frac{4}{3} v^2_\textsc{ta}$ for isotropic materials, see Eq.~\ref{eq:v}. It indicates that $v^\text{-2}_\textsc{k}$ is a direct measure of an {\em apparent density variation} $\delta\rho_\textsc{e}\leq \delta \rho$ related to elastic and frequency dependent viscoelastic processes,
\begin{equation}
\delta\rho_\textsc{e} = \left(v^\text{-2}_\textsc{k} \right) \delta P.  
\label{eq:rhoPvk}
\end{equation}
For a continuous $v_\textsc{k}(P)$ function, $\delta \rho_\textsc{e}$ can always be calculated and does not require knowledge of $\rho$. The cumulative sum of the $\delta \rho_\textsc{e}$ is an apparent density increase $\Delta \rho_\textsc{e}$ as a function of $P$. For an {\em elastic solid}, $\chi_\textsc{e} = \chi_0$ and $\delta\rho_\textsc{e} = \delta\rho$ so that the pressure dependence of the {\em density} can be straightforwardly obtained iteratively using Eq.~\ref{eq:rhoPvk}, starting from the known density $\rho_0$ at a pressure $P_0$~\cite{SCHR1990, ZHA1994, WEIG2012, COAS2014}.

We start our analysis by focusing on the low $P$ part of the compression. Fig.~\ref{fig:Rho} shows the density increase of $v$-SiO$_2$ as a function of the applied pressure for $P\alt\SI{9}{GPa}$. The squares are from volumetric measurements using an accurate strain-gauge technique~\cite{TSIO1998}, thus referring to the static compressibility $\chi_0$. The dashed line in Fig.~\ref{fig:Rho} is the $\rho(P)$ curve calculated using the velocities acquired upon compression and Eq.~\ref{eq:rhoPvk}. The latter is slightly lower than the experimental static values. This is actually explained by the frequency dependence of sound velocities caused by internal friction, \textit{i.e.} viscoelastic effects. Two dissipative processes dominate in glasses at room temperature (See S-II in Supplemental Material): the interaction with relaxing structural entities called {\em defects}~\cite{ANDE1955, JACK1976} and the anharmonic interactions with thermal vibrations~\cite{MARI1971}. Accordingly, sound velocities should vary between a low-frequency (relaxed) value $v_0$ and a high-frequency (unrelaxed) one $v_\infty$. A quantitative description of velocity dispersion exists for $v$-SiO$_2$~\cite{VACH2005} which has been extended to high pressures~\cite{AYRI2011a}. It results in the $v_0(P)$ curve for LA modes shown in the inset in Fig.~\ref{fig:Rho}, significantly lower than the velocities at GHz frequencies (see also Fig.~S2 in Supplemental Material). Using the relaxed velocities $v_\textsc{la,0}(P)$ and $v_\textsc{ta,0}(P)$~\cite{RUFF2010}, a new $\rho(P)$ curve is determined from Eq.~\ref{eq:rhoPvk} which is plotted as a solid line in Fig.~\ref{fig:Rho}. These new $\rho(P)$ variations are appreciably larger than those deduced from uncorrected velocities, and stand very close to the experimental static values. This result highlights the need to take into account the viscoelastic character of the medium to obtain a fair estimate of the EoS. 
 
\begin{figure}
   \centering
   \includegraphics[width=8.6cm]{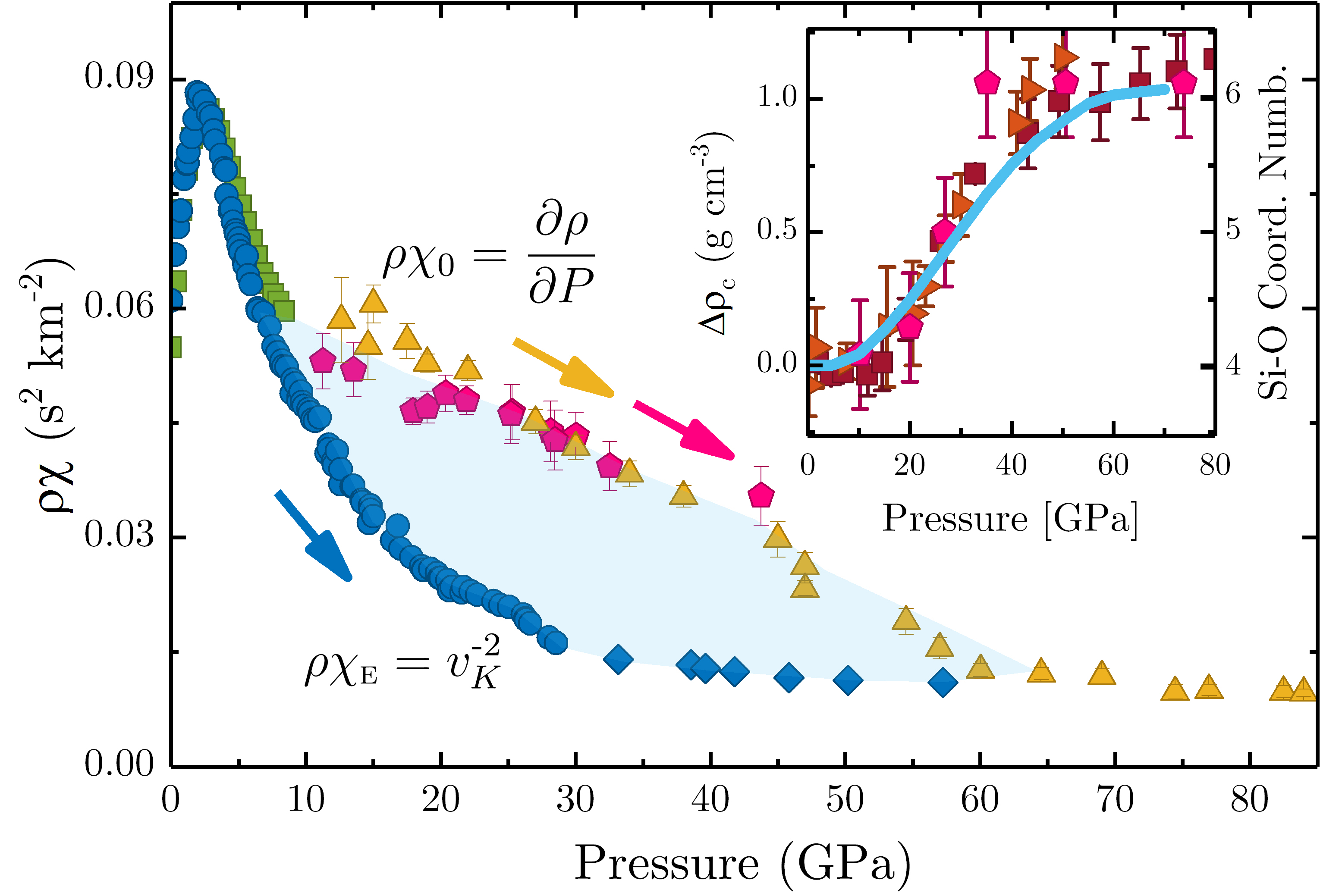} 
   \caption{Static quantity $\rho\chi_0$ from density data (\squadfill[green]~\cite{TSIO1998}, \pentagofill[pink]~\cite{SATO2008}, \trianglepafill[yellow]~\cite{PETI2017}) and $\rho\chi_\textsc{e} = v_\textsc{k}^\text{-2}$ from BLS (\circletfill[blue], \rhombusfill[blue]~\cite{ZHA1994}), upon compression. The blue shaded area is a guide for the eye. Inset: Density increase $\Delta\rho_\textsc{c}$ from Eq.\ref{eq:rho4to6} (solid line) compared to the mean Si$-$O coordination number (\triangleprfill[red]~\cite {BENM2010}, \pentagofill[pink]~\cite{SATO2010} and \squadfill[marron]~\cite{PRES2017}).}
   \label{fig:RhoChi4to6}
\end{figure}

Even corrected from the dynamical effects discussed above, the $v_\textsc{k}$ values obtained for $P \agt$ \SI{10}{GPa} lead however to a compressibility $\chi_\textsc{e}$ which remains much smaller than $\chi_0$. This is shown in Fig.~\ref{fig:RhoChi4to6}, where the pressure dependence of $\rho\chi_\textsc{e} = v_\textsc{k}^\text{-2}$ upon compression (blue circles) is plotted. Our data are completed by those of~\citet{ZHA1994} at the highest $P$ (blue lozenges). These values can be compared to the static ones, $\rho\chi_0 = \partial \rho/\partial P$, calculated using densities from literature~\cite{TSIO1998, SATO2008, PETI2017}, and also shown in Fig.~\ref{fig:RhoChi4to6} (See also S-III and Fig.~S4 in Supplemental Material). Following Fig.~\ref{fig:Rho}, $\rho\chi_\textsc{e}$ and $\rho\chi_0$ superimpose below $\sim\SI{9}{GPa}$. Above, $\rho\chi_\textsc{e}$ decreases much more rapidly than $\rho\chi_0$, their ratio reaching a factor 3 around \SI{30}{GPa}. At $P\simeq\SI{60}{GPa}$ the two curves merge again, indicating that vitreous silica recovers an almost elastic behavior at high $P$.

Two compaction processes can be invoked to explain the difference between the two compressibilities: i) volume non-conservative plastic flow processes which upon decompression would lead to a residual densification. ii) slow relaxation processes {\em appearing frozen in} at the high BLS frequencies, and thus hardening the network, which could result from structural transformations existing in this $P$ range. The two mechanisms can exist side by side and it is \textit{a priori} impossible to estimate them separately from the data plotted in Fig.~\ref{fig:RhoChi4to6}. Let us calculate the compaction $\delta\rho_{\rm c}$ associated with the difference between the two $\rho\chi$ curves for a step $\delta P$ in pressure, 
\begin{equation}
\delta\rho_\textsc{c} = \left(\rho \chi_0 - v_\textsc{k}^\text{-2} \right) \delta P,
\label{eq:rho4to6}
\end{equation}
which relates to the compaction missing in the measured $\rho\chi_\textsc{e}$. The cumulative sum $\Delta\rho_c(P)$ of the latter quantity is reported in the inset in Fig.~\ref{fig:RhoChi4to6} as a solid line. $\Delta\rho_c(P)$ gently starts to increase around \SI{10}{GPa} to eventually end up around \SI{60}{GPa}, having its maximum rate around \SI{30}{GPa}. An increase in density of about$~\SI{1}{g.cm^{-3}}$ is obtained for this contribution. We notice that this value is about twice as high as the known maximum residual densification observed in vitreous silica, \textit{i.e.} \SI{0.45}{g.cm^{-3}}, thus confirming the presence of slow relaxational mechanisms. The mean Si$-$O coordination number as a function of $P$ from recent diffraction experiments~\cite{BENM2010, SATO2010, PRES2017} is also plotted in the inset in Fig.~\ref{fig:RhoChi4to6} using the same relative scale. One observes a remarkable superposition with the calculated $\Delta\rho_{\rm c}(P)$. They almost increase at the same rate, strongly suggesting that $\Delta\rho_{\rm c}(P)$ originates solely from the progressive transformation from the fourfold-coordinated structure to the sixfold one, rather than from plastic flow.

We now turn to the unloading part of the cycles for $P_\text{m}\agt\SI{10}{GPa}$. Comparing the decompression curve from $P_\text{m} = \SI{28.6}{GPa}$ (pink left triangles in Fig.~\ref{fig:vLATA}) to the one from $P_\text{m} = \SI{22.5}{GPa}$ (green pentagons), we first remark that significantly larger velocities are found at high $P$ for the former, while both curves coincide on the low-$P$ side. Re-increasing the pressure from \SI{9}{GPa} (pink right triangles), we observe further that the velocities behave differently from the decompression curve and merge eventually with the initial compression one at $\sim$\SI{19}{GPa}. This suggests an hysteresis phenomenon, illustrated as a pink shaded area in Fig.~\ref{fig:vLATA}. The latter might be due to the fivefold- and sixfold-coordinated Si atoms remaining stable during decompression. In that picture, the progressive structural transformation modifies the elastic properties, both the static and the apparent ones probed by BLS, thus also unveiling the delayed revert transformation through the hysteresis in the sound velocities. From Fig.~\ref{fig:vLATA}), we observe that the return to the tetrahedral structure would then occur between \SIrange{10}{15}{GPa} and more abruptly than the progressive direct transformation upon compression. We can anticipate that volume measurements upon decompression from above \SIrange{25}{30}{GPa} should also reveal similar rapid changes. There are indeed some converging indications of the latter from diffraction experiments~\cite{SATO2018} as well as from \textit{ab initio} molecular dynamics simulations~\cite{RYUO2017}. Further, similar behaviors were also reported for the trigonally coordinated network former $v$-B$_2$O$_3$~\cite{NICH2004, HUAN2008, BRAZ2008, ZEID2014a}. It was proposed that changes in the B$-$O coordination number were gradual upon compression but abrupt at the time of decompression~\cite{NICH2004, HUAN2008}, an interpretation which has however been questioned~\cite{BRAZ2009a}. 

\begin{figure}
  \centering
  \includegraphics[width=8.6cm]{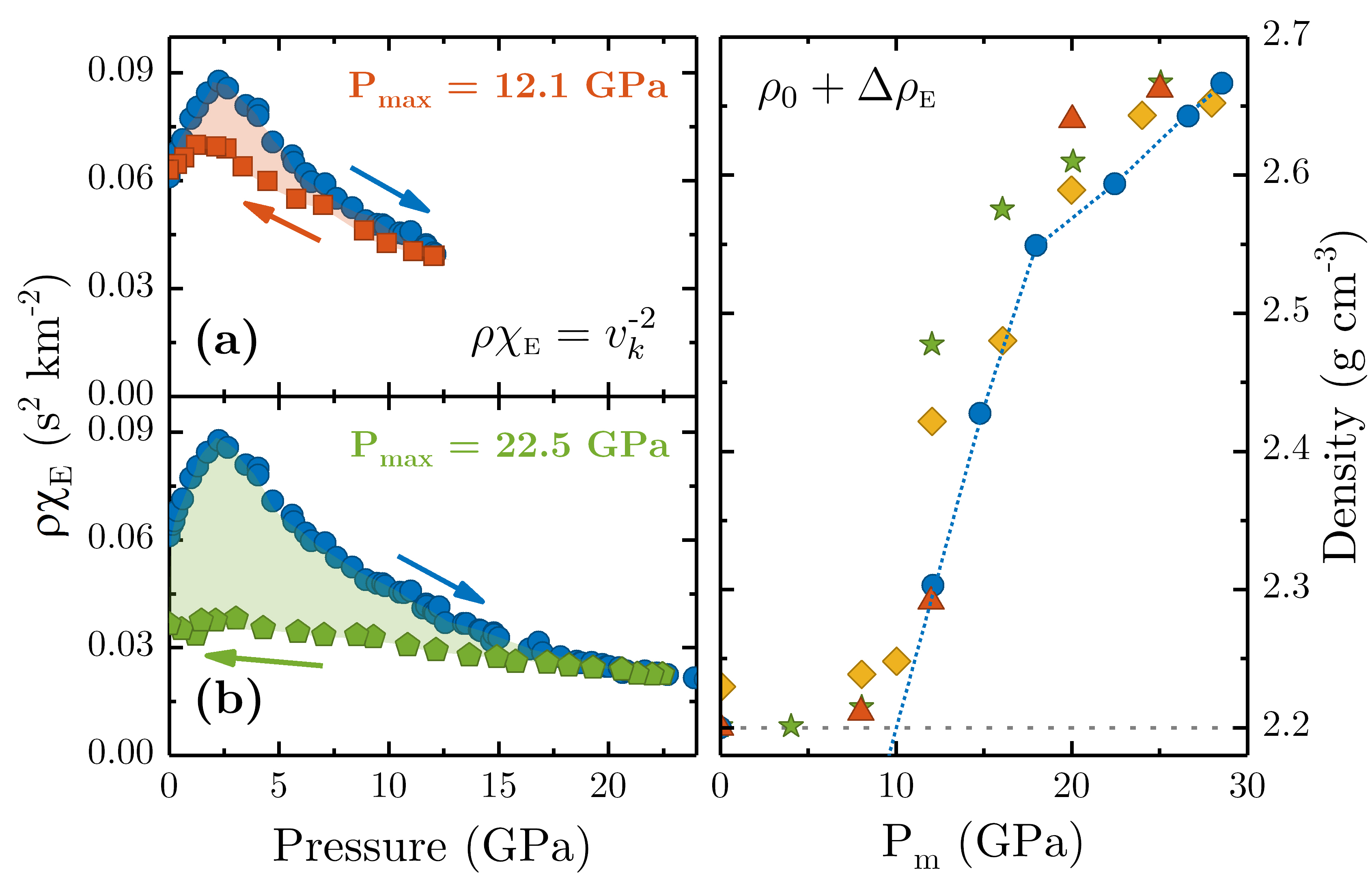}
  \caption{(a) and (b) $\rho\chi_\textsc{e}$ along a compression-decompression cycle for two different $P_\text{m}$ values. The shaded area illustrate the density variations $\Delta\rho_\textsc{e}$. (c) Final density $\rho_0+\Delta\rho_\textsc{e}$ \textit{vs} $P_\text{m}$ (this work \circletfill[blue], \trianglepafill[red]~\cite{ROUX2008}, \starletfill[green]~\cite{SUND2018}, \rhombusfill[yellow]~\cite{LIAN2007}). The dotted line is a guide for the eye.}
  \label{fig:RhochiDens}
\end{figure}

Finally, we address the issue of the residual densification. As shown in Fig.~\ref{fig:vLATA}, the final sound velocity after a complete pressure release is generally higher than that of the pristine sample, reflecting strong modifications in the elastic properties. This final value increases strongly when $P_\text{m}$ goes from \SIrange[range-phrase = ~to~]{12.1}{18}{GPa} and tends to saturate above. Following Eq.~\ref{eq:rhoPvk}, we can calculate the cumulative changes in density $\Delta\rho_\textsc{e}$ along compression-decompression cycles. Figs.~\ref{fig:RhochiDens}a and \ref{fig:RhochiDens}b illustrate the $P$ variations of $\rho\chi_\textsc{e}$ along a cycle for two different $P_\text{m}$. The shaded area, which reflects $\Delta\rho_\textsc{e}$, increases with increasing $P_\text{m}$. The obtained final densities $\rho_0+\Delta\rho_\textsc{e}$ are shown in Fig.~\ref{fig:RhochiDens}(c) as a function of $P_\text{m}$ (blue circles). $\Delta\rho_\textsc{e}$ starts to increase for $P_\text{m} \agt$ \SI{10}{\GPa} until about \SI{20}{\GPa} beyond which it tends to saturate at $\Delta\rho_\textsc{e} \simeq$ \SI{0.45}{g.cm^{-3}}, a value in agreement with the maximum residual densification observed in vitreous silica. To our knowledge, there exists only one set of experimental data for the residual densification as a result of hydrostatic compression at room temperature (red triangles~\cite{ROUX2008}) allowing a comparison with our results at $P_\text{m}=12$, 20 and \SI{25}{GPa}. The latter densities also agree with our outputs. If the $\Delta\rho_\textsc{e}$ had been calculated from the BLS velocities without being corrected for the dynamical dissipative effects, they would have been underestimated by about 10\% (See Fig.~S3 in Supplemental Material). We also note that the hysteresis at high $P$, ascribed to changes in the coordination number, affects the estimate of the residual densification for the highest $P_\text{m}$, however marginally due to the very low compressibility of the highly compacted network (See S-IV and Fig.~S5 in Supplemental Material). Moreover, our results also compare fairly well to estimates of the residual densification based on numerical simulations~\cite{LIAN2007, SUND2018} also plotted in Fig.~\ref{fig:RhochiDens}(c). While showing the same general trend, the latter exhibit slightly larger variations in the initial stages of the densification process. Some other numerical results show qualitatively similar results, albeit with a larger densification rate~\cite{MANT2012}. 

It is remarkable that, even if slow anelastic effects prevent obtaining the EoS from sound velocities using Eq.~\ref{eq:rhoPvk}, the residual densification can be figured out from the latter. This reveals that the underlying process does not relate to volume non-conservative plasticity but rather to continuous changes in the elastic properties of silica along the compression-decompression cycle. It is interesting to note that residual densification, often called {\em permanent densification}, is however reversible at room temperature over periods of several years, showing the metastable character of the residual densification~\cite{POLI1990}. Annealing densified samples at a few hundred degrees above room temperature is sufficient to rapidly recover the pristine density~\cite{MACK1963,RAT1999,GUER2015}. This indicates that irreversible volume non-conservative plastic flow does not occur in hydrostatic compression of silica glass at room temperature, even if this result does not preclude the existence of shear plastic flow in mechanical load testings including shear stress. Residual densification and coordination changes are not two independent processes. The former starts indeed for $P_\text{m}$ above $\sim$\SI{10}{GPa}, in coincidence with the mean Si$-$O coordination number increase and in agreement with the idea that fivefold defects could promote the residual densification~\cite{LIAN2007}. However, residual densification saturates around \SI{20}{GPa} while the mean Si$-$O coordination number reaches 4.5 only~\cite{BENM2010, SATO2010, PRES2017}. To the extent that changes in coordination number disappear at decompression, the residual densification process would rather manifest in the topology of the network \textit{via}, in particular, changes in ring statistics within the structure~\cite{RYUO2017,TAKA2018}.

Summarizing, we show that the large difference between the static compressibility and that extracted from BLS experiments can be used to monitor the pressure window where slow structural rearrangements occur in vitreous silica. The associated compaction rate upon compression is quantified and successfully compared to the Si$-$O average coordination number. Out of this pressure window, vitreous silica behaves almost elastically and the equation of state can be recovered from the sound velocities corrected from dispersion effects, significant at low pressure. Besides, this transformation displays a significant hysteresis. Finally, we find out that the large difference observed in the sound velocity data along the loading-unloading path does account for the residual densification of the recovered silica samples. This reveals that the underlying process does not relate to volume non-conservative plasticity but rather to continuous changes in the elastic properties.

More generally, this work also shows that complete pressure cycles certainly deserve more attention and that precise density {\em and} sound velocity measurements, when combined, give fundamental insights into pressure-induced transformations in glasses. That would be definitely interesting to address the case of $v$-B$_2$O$_3$ or $v$-GeO$_2$ for example. For the latter, there are indications that polyamorphism with coordination numbers higher than six occurs above \SI{30}{GPa}~\cite{BRAZ2011,KONO2016}, which would justify similar work. Concerning $v$-B$_2$O$_3$, such approach should help in solving the controversy existing about the rapid changes in the structure and the elastic properties upon decompression. 

This work was supported by the French National Research Agency programs MECASIL ANR-12-BS04-0004-03 and PIPOG  ANR-17-CE30-0009.


\begin{thebibliography}{54}%
\makeatletter
\providecommand \@ifxundefined [1]{%
 \@ifx{#1\undefined}
}%
\providecommand \@ifnum [1]{%
 \ifnum #1\expandafter \@firstoftwo
 \else \expandafter \@secondoftwo
 \fi
}%
\providecommand \@ifx [1]{%
 \ifx #1\expandafter \@firstoftwo
 \else \expandafter \@secondoftwo
 \fi
}%
\providecommand \natexlab [1]{#1}%
\providecommand \enquote  [1]{``#1''}%
\providecommand \bibnamefont  [1]{#1}%
\providecommand \bibfnamefont [1]{#1}%
\providecommand \citenamefont [1]{#1}%
\providecommand \href@noop [0]{\@secondoftwo}%
\providecommand \href [0]{\begingroup \@sanitize@url \@href}%
\providecommand \@href[1]{\@@startlink{#1}\@@href}%
\providecommand \@@href[1]{\endgroup#1\@@endlink}%
\providecommand \@sanitize@url [0]{\catcode `\\12\catcode `\$12\catcode
  `\&12\catcode `\#12\catcode `\^12\catcode `\_12\catcode `\%12\relax}%
\providecommand \@@startlink[1]{}%
\providecommand \@@endlink[0]{}%
\providecommand \url  [0]{\begingroup\@sanitize@url \@url }%
\providecommand \@url [1]{\endgroup\@href {#1}{\urlprefix }}%
\providecommand \urlprefix  [0]{URL }%
\providecommand \Eprint [0]{\href }%
\providecommand \doibase [0]{http://dx.doi.org/}%
\providecommand \selectlanguage [0]{\@gobble}%
\providecommand \bibinfo  [0]{\@secondoftwo}%
\providecommand \bibfield  [0]{\@secondoftwo}%
\providecommand \translation [1]{[#1]}%
\providecommand \BibitemOpen [0]{}%
\providecommand \bibitemStop [0]{}%
\providecommand \bibitemNoStop [0]{.\EOS\space}%
\providecommand \EOS [0]{\spacefactor3000\relax}%
\providecommand \BibitemShut  [1]{\csname bibitem#1\endcsname}%
\let\auto@bib@innerbib\@empty
\bibitem [{\citenamefont {McMillan}(2002)}]{MCMI2002}%
  \BibitemOpen
  \bibfield  {author} {\bibinfo {author} {\bibfnamefont {P.}~\bibnamefont
  {McMillan}},\ }\href {\doibase 10.1038/nmat716} {\bibfield  {journal}
  {\bibinfo  {journal} {Nat. Mater.}\ }\textbf {\bibinfo {volume} {1}},\
  \bibinfo {pages} {19} (\bibinfo {year} {2002})}\BibitemShut {NoStop}%
\bibitem [{\citenamefont {Brazhkin}(2007)}]{BRAZ2007}%
  \BibitemOpen
  \bibfield  {author} {\bibinfo {author} {\bibfnamefont {V.}~\bibnamefont
  {Brazhkin}},\ }\href {\doibase 10.1080/08957950701546956} {\bibfield
  {journal} {\bibinfo  {journal} {High Pressure Res.}\ }\textbf {\bibinfo
  {volume} {27}},\ \bibinfo {pages} {333} (\bibinfo {year} {2007})}\BibitemShut
  {NoStop}%
\bibitem [{\citenamefont {Sanloup}\ \emph {et~al.}(2013)\citenamefont
  {Sanloup}, \citenamefont {Drewitt}, \citenamefont {Konopkova}, \citenamefont
  {Dalladay-Simpson}, \citenamefont {Morton}, \citenamefont {Rai},
  \citenamefont {van Westrenen},\ and\ \citenamefont {Morgenroth}}]{SANL2013}%
  \BibitemOpen
  \bibfield  {author} {\bibinfo {author} {\bibfnamefont {C.}~\bibnamefont
  {Sanloup}}, \bibinfo {author} {\bibfnamefont {J.~W.~E.}\ \bibnamefont
  {Drewitt}}, \bibinfo {author} {\bibfnamefont {Z.}~\bibnamefont {Konopkova}},
  \bibinfo {author} {\bibfnamefont {P.}~\bibnamefont {Dalladay-Simpson}},
  \bibinfo {author} {\bibfnamefont {D.~M.}\ \bibnamefont {Morton}}, \bibinfo
  {author} {\bibfnamefont {N.}~\bibnamefont {Rai}}, \bibinfo {author}
  {\bibfnamefont {W.}~\bibnamefont {van Westrenen}}, \ and\ \bibinfo {author}
  {\bibfnamefont {W.}~\bibnamefont {Morgenroth}},\ }\href {\doibase
  10.1038/nature12668} {\bibfield  {journal} {\bibinfo  {journal} {Nature}\
  }\textbf {\bibinfo {volume} {503}},\ \bibinfo {pages} {104} (\bibinfo {year}
  {2013})}\BibitemShut {NoStop}%
\bibitem [{\citenamefont {Millot}\ \emph {et~al.}(2015)\citenamefont {Millot},
  \citenamefont {Dubrovinskaia}, \citenamefont {{\v C}ernok}, \citenamefont
  {Blaha}, \citenamefont {Dubrovinsky}, \citenamefont {Braun}, \citenamefont
  {Celliers}, \citenamefont {Collins}, \citenamefont {Eggert},\ and\
  \citenamefont {Jeanloz}}]{MILL2015}%
  \BibitemOpen
  \bibfield  {author} {\bibinfo {author} {\bibfnamefont {M.}~\bibnamefont
  {Millot}}, \bibinfo {author} {\bibfnamefont {N.}~\bibnamefont
  {Dubrovinskaia}}, \bibinfo {author} {\bibfnamefont {A.}~\bibnamefont {{\v
  C}ernok}}, \bibinfo {author} {\bibfnamefont {S.}~\bibnamefont {Blaha}},
  \bibinfo {author} {\bibfnamefont {L.}~\bibnamefont {Dubrovinsky}}, \bibinfo
  {author} {\bibfnamefont {D.~G.}\ \bibnamefont {Braun}}, \bibinfo {author}
  {\bibfnamefont {P.~M.}\ \bibnamefont {Celliers}}, \bibinfo {author}
  {\bibfnamefont {G.~W.}\ \bibnamefont {Collins}}, \bibinfo {author}
  {\bibfnamefont {J.~H.}\ \bibnamefont {Eggert}}, \ and\ \bibinfo {author}
  {\bibfnamefont {R.}~\bibnamefont {Jeanloz}},\ }\href {\doibase
  10.1126/science.1261507} {\bibfield  {journal} {\bibinfo  {journal}
  {Science}\ }\textbf {\bibinfo {volume} {347}},\ \bibinfo {pages} {418}
  (\bibinfo {year} {2015})}\BibitemShut {NoStop}%
\bibitem [{\citenamefont {Meade}\ and\ \citenamefont
  {Jeanloz}(1987)}]{MEAD1987}%
  \BibitemOpen
  \bibfield  {author} {\bibinfo {author} {\bibfnamefont {C.}~\bibnamefont
  {Meade}}\ and\ \bibinfo {author} {\bibfnamefont {R.}~\bibnamefont
  {Jeanloz}},\ }\href {\doibase 10.1103/PhysRevB.35.236} {\bibfield  {journal}
  {\bibinfo  {journal} {Phys. Rev. B}\ }\textbf {\bibinfo {volume} {35}},\
  \bibinfo {pages} {236} (\bibinfo {year} {1987})}\BibitemShut {NoStop}%
\bibitem [{\citenamefont {Tsiok}\ \emph {et~al.}(1998)\citenamefont {Tsiok},
  \citenamefont {Brazhkin}, \citenamefont {Lyapin},\ and\ \citenamefont
  {Khvostantsev}}]{TSIO1998}%
  \BibitemOpen
  \bibfield  {author} {\bibinfo {author} {\bibfnamefont {O.~B.}\ \bibnamefont
  {Tsiok}}, \bibinfo {author} {\bibfnamefont {V.~V.}\ \bibnamefont {Brazhkin}},
  \bibinfo {author} {\bibfnamefont {A.~G.}\ \bibnamefont {Lyapin}}, \ and\
  \bibinfo {author} {\bibfnamefont {L.~G.}\ \bibnamefont {Khvostantsev}},\
  }\href {\doibase 10.1103/PhysRevLett.80.999} {\bibfield  {journal} {\bibinfo
  {journal} {Phys. Rev. Lett.}\ }\textbf {\bibinfo {volume} {80}},\ \bibinfo
  {pages} {999} (\bibinfo {year} {1998})}\BibitemShut {NoStop}%
\bibitem [{\citenamefont {Polian}\ and\ \citenamefont
  {Grimsditch}(1993)}]{POLI1993}%
  \BibitemOpen
  \bibfield  {author} {\bibinfo {author} {\bibfnamefont {A.}~\bibnamefont
  {Polian}}\ and\ \bibinfo {author} {\bibfnamefont {M.}~\bibnamefont
  {Grimsditch}},\ }\href {https://link.aps.org/doi/10.1103/PhysRevB.47.13979}
  {\bibfield  {journal} {\bibinfo  {journal} {Phys. Rev. B}\ }\textbf {\bibinfo
  {volume} {47}},\ \bibinfo {pages} {13979} (\bibinfo {year}
  {1993})}\BibitemShut {NoStop}%
\bibitem [{\citenamefont {Sato}\ and\ \citenamefont
  {Funamori}(2008)}]{SATO2008}%
  \BibitemOpen
  \bibfield  {author} {\bibinfo {author} {\bibfnamefont {T.}~\bibnamefont
  {Sato}}\ and\ \bibinfo {author} {\bibfnamefont {N.}~\bibnamefont
  {Funamori}},\ }\href
  {https://link.aps.org/doi/10.1103/PhysRevLett.101.255502} {\bibfield
  {journal} {\bibinfo  {journal} {Phys. Rev. Lett.}\ }\textbf {\bibinfo
  {volume} {101}},\ \bibinfo {pages} {255502} (\bibinfo {year}
  {2008})}\BibitemShut {NoStop}%
\bibitem [{\citenamefont {Hemley}\ \emph {et~al.}(1986)\citenamefont {Hemley},
  \citenamefont {Mao}, \citenamefont {Bell},\ and\ \citenamefont
  {Mysen}}]{HEML1986}%
  \BibitemOpen
  \bibfield  {author} {\bibinfo {author} {\bibfnamefont {R.~J.}\ \bibnamefont
  {Hemley}}, \bibinfo {author} {\bibfnamefont {H.~K.}\ \bibnamefont {Mao}},
  \bibinfo {author} {\bibfnamefont {P.~M.}\ \bibnamefont {Bell}}, \ and\
  \bibinfo {author} {\bibfnamefont {B.~O.}\ \bibnamefont {Mysen}},\ }\href
  {\doibase 10.1103/PhysRevLett.57.747} {\bibfield  {journal} {\bibinfo
  {journal} {Phys. Rev. Lett.}\ }\textbf {\bibinfo {volume} {57}},\ \bibinfo
  {pages} {747} (\bibinfo {year} {1986})}\BibitemShut {NoStop}%
\bibitem [{\citenamefont {Williams}\ and\ \citenamefont
  {Jeanloz}(1988)}]{WILL1988}%
  \BibitemOpen
  \bibfield  {author} {\bibinfo {author} {\bibfnamefont {Q.}~\bibnamefont
  {Williams}}\ and\ \bibinfo {author} {\bibfnamefont {R.}~\bibnamefont
  {Jeanloz}},\ }\href {http://science.sciencemag.org/content/239/4842/902}
  {\bibfield  {journal} {\bibinfo  {journal} {Science}\ }\textbf {\bibinfo
  {volume} {239}},\ \bibinfo {pages} {902} (\bibinfo {year}
  {1988})}\BibitemShut {NoStop}%
\bibitem [{\citenamefont {Susman}\ \emph {et~al.}(1991)\citenamefont {Susman},
  \citenamefont {Volin}, \citenamefont {Price}, \citenamefont {Grimsditch},
  \citenamefont {Rino}, \citenamefont {Kalia}, \citenamefont {Vashishta},
  \citenamefont {Gwanmesia}, \citenamefont {Wang},\ and\ \citenamefont
  {Liebermann}}]{SUSM1991}%
  \BibitemOpen
  \bibfield  {author} {\bibinfo {author} {\bibfnamefont {S.}~\bibnamefont
  {Susman}}, \bibinfo {author} {\bibfnamefont {K.~J.}\ \bibnamefont {Volin}},
  \bibinfo {author} {\bibfnamefont {D.~L.}\ \bibnamefont {Price}}, \bibinfo
  {author} {\bibfnamefont {M.}~\bibnamefont {Grimsditch}}, \bibinfo {author}
  {\bibfnamefont {J.~P.}\ \bibnamefont {Rino}}, \bibinfo {author}
  {\bibfnamefont {R.~K.}\ \bibnamefont {Kalia}}, \bibinfo {author}
  {\bibfnamefont {P.}~\bibnamefont {Vashishta}}, \bibinfo {author}
  {\bibfnamefont {G.}~\bibnamefont {Gwanmesia}}, \bibinfo {author}
  {\bibfnamefont {Y.}~\bibnamefont {Wang}}, \ and\ \bibinfo {author}
  {\bibfnamefont {R.~C.}\ \bibnamefont {Liebermann}},\ }\href
  {https://link.aps.org/doi/10.1103/PhysRevB.43.1194} {\bibfield  {journal}
  {\bibinfo  {journal} {Phys. Rev. B}\ }\textbf {\bibinfo {volume} {43}},\
  \bibinfo {pages} {1194} (\bibinfo {year} {1991})}\BibitemShut {NoStop}%
\bibitem [{\citenamefont {Meade}\ \emph {et~al.}(1992)\citenamefont {Meade},
  \citenamefont {Hemley},\ and\ \citenamefont {Mao}}]{MEAD1992}%
  \BibitemOpen
  \bibfield  {author} {\bibinfo {author} {\bibfnamefont {C.}~\bibnamefont
  {Meade}}, \bibinfo {author} {\bibfnamefont {R.~J.}\ \bibnamefont {Hemley}}, \
  and\ \bibinfo {author} {\bibfnamefont {H.~K.}\ \bibnamefont {Mao}},\ }\href
  {\doibase 10.1103/PhysRevLett.69.1387} {\bibfield  {journal} {\bibinfo
  {journal} {Phys. Rev. Lett.}\ }\textbf {\bibinfo {volume} {69}},\ \bibinfo
  {pages} {1387} (\bibinfo {year} {1992})}\BibitemShut {NoStop}%
\bibitem [{\citenamefont {Zha}\ \emph {et~al.}(1994)\citenamefont {Zha},
  \citenamefont {Hemley}, \citenamefont {Mao}, \citenamefont {Duffy},\ and\
  \citenamefont {Meade}}]{ZHA1994}%
  \BibitemOpen
  \bibfield  {author} {\bibinfo {author} {\bibfnamefont {C.-s.}\ \bibnamefont
  {Zha}}, \bibinfo {author} {\bibfnamefont {R.~J.}\ \bibnamefont {Hemley}},
  \bibinfo {author} {\bibfnamefont {H.-k.}\ \bibnamefont {Mao}}, \bibinfo
  {author} {\bibfnamefont {T.~S.}\ \bibnamefont {Duffy}}, \ and\ \bibinfo
  {author} {\bibfnamefont {C.}~\bibnamefont {Meade}},\ }\href {\doibase
  10.1103/PhysRevB.50.13105} {\bibfield  {journal} {\bibinfo  {journal} {Phys.
  Rev. B}\ }\textbf {\bibinfo {volume} {50}},\ \bibinfo {pages} {13105}
  (\bibinfo {year} {1994})}\BibitemShut {NoStop}%
\bibitem [{\citenamefont {Brazhkin}(2009)}]{BRAZ2009b}%
  \BibitemOpen
  \bibfield  {author} {\bibinfo {author} {\bibfnamefont {V.~V.}\ \bibnamefont
  {Brazhkin}},\ }\href
  {https://link.aps.org/doi/10.1103/PhysRevLett.102.209603} {\bibfield
  {journal} {\bibinfo  {journal} {Phys. Rev. Lett.}\ }\textbf {\bibinfo
  {volume} {102}},\ \bibinfo {pages} {209603} (\bibinfo {year}
  {2009})}\BibitemShut {NoStop}%
\bibitem [{\citenamefont {Benmore}\ \emph {et~al.}(2010)\citenamefont
  {Benmore}, \citenamefont {Soignard}, \citenamefont {Amin}, \citenamefont
  {Guthrie}, \citenamefont {Shastri}, \citenamefont {Lee},\ and\ \citenamefont
  {Yarger}}]{BENM2010}%
  \BibitemOpen
  \bibfield  {author} {\bibinfo {author} {\bibfnamefont {C.~J.}\ \bibnamefont
  {Benmore}}, \bibinfo {author} {\bibfnamefont {E.}~\bibnamefont {Soignard}},
  \bibinfo {author} {\bibfnamefont {S.~A.}\ \bibnamefont {Amin}}, \bibinfo
  {author} {\bibfnamefont {M.}~\bibnamefont {Guthrie}}, \bibinfo {author}
  {\bibfnamefont {S.~D.}\ \bibnamefont {Shastri}}, \bibinfo {author}
  {\bibfnamefont {P.~L.}\ \bibnamefont {Lee}}, \ and\ \bibinfo {author}
  {\bibfnamefont {J.~L.}\ \bibnamefont {Yarger}},\ }\href
  {https://link.aps.org/doi/10.1103/PhysRevB.81.054105} {\bibfield  {journal}
  {\bibinfo  {journal} {Phys. Rev. B}\ }\textbf {\bibinfo {volume} {81}},\
  \bibinfo {pages} {054105} (\bibinfo {year} {2010})}\BibitemShut {NoStop}%
\bibitem [{\citenamefont {Sato}\ and\ \citenamefont
  {Funamori}(2010)}]{SATO2010}%
  \BibitemOpen
  \bibfield  {author} {\bibinfo {author} {\bibfnamefont {T.}~\bibnamefont
  {Sato}}\ and\ \bibinfo {author} {\bibfnamefont {N.}~\bibnamefont
  {Funamori}},\ }\href {https://link.aps.org/doi/10.1103/PhysRevB.82.184102}
  {\bibfield  {journal} {\bibinfo  {journal} {Phys. Rev. B}\ }\textbf {\bibinfo
  {volume} {82}},\ \bibinfo {pages} {184102} (\bibinfo {year}
  {2010})}\BibitemShut {NoStop}%
\bibitem [{\citenamefont {Zeidler}\ \emph
  {et~al.}(2014{\natexlab{a}})\citenamefont {Zeidler}, \citenamefont {Wezka},
  \citenamefont {Rowlands}, \citenamefont {Whittaker}, \citenamefont {Salmon},
  \citenamefont {Polidori}, \citenamefont {Drewitt}, \citenamefont {Klotz},
  \citenamefont {Fischer}, \citenamefont {Wilding}, \citenamefont {Bull},
  \citenamefont {Tucker},\ and\ \citenamefont {Wilson}}]{ZEID2014b}%
  \BibitemOpen
  \bibfield  {author} {\bibinfo {author} {\bibfnamefont {A.}~\bibnamefont
  {Zeidler}}, \bibinfo {author} {\bibfnamefont {K.}~\bibnamefont {Wezka}},
  \bibinfo {author} {\bibfnamefont {R.~F.}\ \bibnamefont {Rowlands}}, \bibinfo
  {author} {\bibfnamefont {D.~A.~J.}\ \bibnamefont {Whittaker}}, \bibinfo
  {author} {\bibfnamefont {P.~S.}\ \bibnamefont {Salmon}}, \bibinfo {author}
  {\bibfnamefont {A.}~\bibnamefont {Polidori}}, \bibinfo {author}
  {\bibfnamefont {J.~W.~E.}\ \bibnamefont {Drewitt}}, \bibinfo {author}
  {\bibfnamefont {S.}~\bibnamefont {Klotz}}, \bibinfo {author} {\bibfnamefont
  {H.~E.}\ \bibnamefont {Fischer}}, \bibinfo {author} {\bibfnamefont {M.~C.}\
  \bibnamefont {Wilding}}, \bibinfo {author} {\bibfnamefont {C.~L.}\
  \bibnamefont {Bull}}, \bibinfo {author} {\bibfnamefont {M.~G.}\ \bibnamefont
  {Tucker}}, \ and\ \bibinfo {author} {\bibfnamefont {M.}~\bibnamefont
  {Wilson}},\ }\href {\doibase 10.1103/PhysRevLett.113.135501} {\bibfield
  {journal} {\bibinfo  {journal} {Phys. Rev. Lett.}\ }\textbf {\bibinfo
  {volume} {113}},\ \bibinfo {pages} {135501} (\bibinfo {year}
  {2014}{\natexlab{a}})}\BibitemShut {NoStop}%
\bibitem [{\citenamefont {Trease}\ \emph {et~al.}(2017)\citenamefont {Trease},
  \citenamefont {Clark}, \citenamefont {Grandinetti}, \citenamefont
  {Stebbins},\ and\ \citenamefont {Sen}}]{TREA2017}%
  \BibitemOpen
  \bibfield  {author} {\bibinfo {author} {\bibfnamefont {N.~M.}\ \bibnamefont
  {Trease}}, \bibinfo {author} {\bibfnamefont {T.~M.}\ \bibnamefont {Clark}},
  \bibinfo {author} {\bibfnamefont {P.~J.}\ \bibnamefont {Grandinetti}},
  \bibinfo {author} {\bibfnamefont {J.~F.}\ \bibnamefont {Stebbins}}, \ and\
  \bibinfo {author} {\bibfnamefont {S.}~\bibnamefont {Sen}},\ }\href {\doibase
  10.1063/1.4983041} {\bibfield  {journal} {\bibinfo  {journal} {J. Chem.
  Phys.}\ }\textbf {\bibinfo {volume} {146}},\ \bibinfo {pages} {184505}
  (\bibinfo {year} {2017})}\BibitemShut {NoStop}%
\bibitem [{\citenamefont {Prescher}\ \emph {et~al.}(2017)\citenamefont
  {Prescher}, \citenamefont {Prakapenka}, \citenamefont {Stefanski},
  \citenamefont {Jahn}, \citenamefont {Skinner},\ and\ \citenamefont
  {Wang}}]{PRES2017}%
  \BibitemOpen
  \bibfield  {author} {\bibinfo {author} {\bibfnamefont {C.}~\bibnamefont
  {Prescher}}, \bibinfo {author} {\bibfnamefont {V.~B.}\ \bibnamefont
  {Prakapenka}}, \bibinfo {author} {\bibfnamefont {J.}~\bibnamefont
  {Stefanski}}, \bibinfo {author} {\bibfnamefont {S.}~\bibnamefont {Jahn}},
  \bibinfo {author} {\bibfnamefont {L.~B.}\ \bibnamefont {Skinner}}, \ and\
  \bibinfo {author} {\bibfnamefont {Y.}~\bibnamefont {Wang}},\ }\href
  {http://www.pnas.org/content/114/38/10041} {\bibfield  {journal} {\bibinfo
  {journal} {Proc. Natl. Acad. Sci.}\ }\textbf {\bibinfo {volume} {114}},\
  \bibinfo {pages} {10041} (\bibinfo {year} {2017})}\BibitemShut {NoStop}%
\bibitem [{\citenamefont {Kondo}\ \emph {et~al.}(1981)\citenamefont {Kondo},
  \citenamefont {Lio},\ and\ \citenamefont {Sawaoka}}]{KOND1981}%
  \BibitemOpen
  \bibfield  {author} {\bibinfo {author} {\bibfnamefont {K.}~\bibnamefont
  {Kondo}}, \bibinfo {author} {\bibfnamefont {S.}~\bibnamefont {Lio}}, \ and\
  \bibinfo {author} {\bibfnamefont {A.}~\bibnamefont {Sawaoka}},\ }\href
  {\doibase 10.1063/1.329012} {\bibfield  {journal} {\bibinfo  {journal} {J.
  Appl. Phys.}\ }\textbf {\bibinfo {volume} {52}},\ \bibinfo {pages} {2826}
  (\bibinfo {year} {1981})}\BibitemShut {NoStop}%
\bibitem [{\citenamefont {Schroeder}\ \emph {et~al.}(1990)\citenamefont
  {Schroeder}, \citenamefont {Bilodeau},\ and\ \citenamefont
  {Zhao}}]{SCHR1990}%
  \BibitemOpen
  \bibfield  {author} {\bibinfo {author} {\bibfnamefont {J.}~\bibnamefont
  {Schroeder}}, \bibinfo {author} {\bibfnamefont {T.~G.}\ \bibnamefont
  {Bilodeau}}, \ and\ \bibinfo {author} {\bibfnamefont {X.-S.}\ \bibnamefont
  {Zhao}},\ }\href {\doibase 10.1080/08957959008246178} {\bibfield  {journal}
  {\bibinfo  {journal} {High Pressure Res.}\ }\textbf {\bibinfo {volume} {4}},\
  \bibinfo {pages} {531} (\bibinfo {year} {1990})}\BibitemShut {NoStop}%
\bibitem [{\citenamefont {Weigel}\ \emph {et~al.}(2012)\citenamefont {Weigel},
  \citenamefont {Polian}, \citenamefont {Kint}, \citenamefont {Ruffl\'e},
  \citenamefont {Foret},\ and\ \citenamefont {Vacher}}]{WEIG2012}%
  \BibitemOpen
  \bibfield  {author} {\bibinfo {author} {\bibfnamefont {C.}~\bibnamefont
  {Weigel}}, \bibinfo {author} {\bibfnamefont {A.}~\bibnamefont {Polian}},
  \bibinfo {author} {\bibfnamefont {M.}~\bibnamefont {Kint}}, \bibinfo {author}
  {\bibfnamefont {B.}~\bibnamefont {Ruffl\'e}}, \bibinfo {author}
  {\bibfnamefont {M.}~\bibnamefont {Foret}}, \ and\ \bibinfo {author}
  {\bibfnamefont {R.}~\bibnamefont {Vacher}},\ }\href {\doibase
  10.1103/PhysRevLett.109.245504} {\bibfield  {journal} {\bibinfo  {journal}
  {Phys. Rev. Lett.}\ }\textbf {\bibinfo {volume} {109}},\ \bibinfo {pages}
  {245504} (\bibinfo {year} {2012})}\BibitemShut {NoStop}%
\bibitem [{\citenamefont {Coasne}\ \emph {et~al.}(2014)\citenamefont {Coasne},
  \citenamefont {Weigel}, \citenamefont {Polian}, \citenamefont {Kint},
  \citenamefont {Rouquette}, \citenamefont {Haines}, \citenamefont {Foret},
  \citenamefont {Vacher},\ and\ \citenamefont {Rufflé}}]{COAS2014}%
  \BibitemOpen
  \bibfield  {author} {\bibinfo {author} {\bibfnamefont {B.}~\bibnamefont
  {Coasne}}, \bibinfo {author} {\bibfnamefont {C.}~\bibnamefont {Weigel}},
  \bibinfo {author} {\bibfnamefont {A.}~\bibnamefont {Polian}}, \bibinfo
  {author} {\bibfnamefont {M.}~\bibnamefont {Kint}}, \bibinfo {author}
  {\bibfnamefont {J.}~\bibnamefont {Rouquette}}, \bibinfo {author}
  {\bibfnamefont {J.}~\bibnamefont {Haines}}, \bibinfo {author} {\bibfnamefont
  {M.}~\bibnamefont {Foret}}, \bibinfo {author} {\bibfnamefont
  {R.}~\bibnamefont {Vacher}}, \ and\ \bibinfo {author} {\bibfnamefont
  {B.}~\bibnamefont {Rufflé}},\ }\href {https://doi.org/10.1021/jp5094383}
  {\bibfield  {journal} {\bibinfo  {journal} {J. Phys. Chem. B}\ }\textbf
  {\bibinfo {volume} {118}},\ \bibinfo {pages} {14519} (\bibinfo {year}
  {2014})}\BibitemShut {NoStop}%
\bibitem [{\citenamefont {Grimsditch}(1984)}]{GRIM1984}%
  \BibitemOpen
  \bibfield  {author} {\bibinfo {author} {\bibfnamefont {M.}~\bibnamefont
  {Grimsditch}},\ }\href {\doibase 10.1103/PhysRevLett.52.2379} {\bibfield
  {journal} {\bibinfo  {journal} {Phys. Rev. Lett.}\ }\textbf {\bibinfo
  {volume} {52}},\ \bibinfo {pages} {2379} (\bibinfo {year}
  {1984})}\BibitemShut {NoStop}%
\bibitem [{\citenamefont {Polian}\ and\ \citenamefont
  {Grimsditch}(1990)}]{POLI1990}%
  \BibitemOpen
  \bibfield  {author} {\bibinfo {author} {\bibfnamefont {A.}~\bibnamefont
  {Polian}}\ and\ \bibinfo {author} {\bibfnamefont {M.}~\bibnamefont
  {Grimsditch}},\ }\href {\doibase 10.1103/PhysRevB.41.6086} {\bibfield
  {journal} {\bibinfo  {journal} {Phys. Rev. B}\ }\textbf {\bibinfo {volume}
  {41}},\ \bibinfo {pages} {6086} (\bibinfo {year} {1990})}\BibitemShut
  {NoStop}%
\bibitem [{\citenamefont {Rouxel}\ \emph {et~al.}(2010)\citenamefont {Rouxel},
  \citenamefont {Ji}, \citenamefont {Guin}, \citenamefont {Augereau},\ and\
  \citenamefont {Ruffl\'e}}]{ROUX2010}%
  \BibitemOpen
  \bibfield  {author} {\bibinfo {author} {\bibfnamefont {T.}~\bibnamefont
  {Rouxel}}, \bibinfo {author} {\bibfnamefont {H.}~\bibnamefont {Ji}}, \bibinfo
  {author} {\bibfnamefont {J.~P.}\ \bibnamefont {Guin}}, \bibinfo {author}
  {\bibfnamefont {F.}~\bibnamefont {Augereau}}, \ and\ \bibinfo {author}
  {\bibfnamefont {B.}~\bibnamefont {Ruffl\'e}},\ }\href {\doibase
  10.1063/1.3407559} {\bibfield  {journal} {\bibinfo  {journal} {J. Appl.
  Phys.}\ }\textbf {\bibinfo {volume} {107}},\ \bibinfo {pages} {094903}
  (\bibinfo {year} {2010})}\BibitemShut {NoStop}%
\bibitem [{\citenamefont {Brazhkin}\ \emph {et~al.}(2008)\citenamefont
  {Brazhkin}, \citenamefont {Katayama}, \citenamefont {Trachenko},
  \citenamefont {Tsiok}, \citenamefont {Lyapin}, \citenamefont {Artacho},
  \citenamefont {Dove}, \citenamefont {Ferlat}, \citenamefont {Inamura},\ and\
  \citenamefont {Saitoh}}]{BRAZ2008}%
  \BibitemOpen
  \bibfield  {author} {\bibinfo {author} {\bibfnamefont {V.~V.}\ \bibnamefont
  {Brazhkin}}, \bibinfo {author} {\bibfnamefont {Y.}~\bibnamefont {Katayama}},
  \bibinfo {author} {\bibfnamefont {K.}~\bibnamefont {Trachenko}}, \bibinfo
  {author} {\bibfnamefont {O.~B.}\ \bibnamefont {Tsiok}}, \bibinfo {author}
  {\bibfnamefont {A.~G.}\ \bibnamefont {Lyapin}}, \bibinfo {author}
  {\bibfnamefont {E.}~\bibnamefont {Artacho}}, \bibinfo {author} {\bibfnamefont
  {M.}~\bibnamefont {Dove}}, \bibinfo {author} {\bibfnamefont {G.}~\bibnamefont
  {Ferlat}}, \bibinfo {author} {\bibfnamefont {Y.}~\bibnamefont {Inamura}}, \
  and\ \bibinfo {author} {\bibfnamefont {H.}~\bibnamefont {Saitoh}},\ }\href
  {https://link.aps.org/doi/10.1103/PhysRevLett.101.035702} {\bibfield
  {journal} {\bibinfo  {journal} {Phys. Rev. Lett.}\ }\textbf {\bibinfo
  {volume} {101}},\ \bibinfo {pages} {035702} (\bibinfo {year}
  {2008})}\BibitemShut {NoStop}%
\bibitem [{\citenamefont {Brazhkin}\ \emph {et~al.}(2016)\citenamefont
  {Brazhkin}, \citenamefont {Bychkov},\ and\ \citenamefont {Tsiok}}]{BRAZ2016}%
  \BibitemOpen
  \bibfield  {author} {\bibinfo {author} {\bibfnamefont {V.~V.}\ \bibnamefont
  {Brazhkin}}, \bibinfo {author} {\bibfnamefont {E.}~\bibnamefont {Bychkov}}, \
  and\ \bibinfo {author} {\bibfnamefont {O.~B.}\ \bibnamefont {Tsiok}},\ }\href
  {\doibase 10.1021/acs.jpcb.5b10559} {\bibfield  {journal} {\bibinfo
  {journal} {J. Phys. Chem. B}\ }\textbf {\bibinfo {volume} {120}},\ \bibinfo
  {pages} {358} (\bibinfo {year} {2016})}\BibitemShut {NoStop}%
\bibitem [{\citenamefont {Chervin}\ \emph {et~al.}(2001)\citenamefont
  {Chervin}, \citenamefont {Canny},\ and\ \citenamefont
  {Mancinelli}}]{CHER2001}%
  \BibitemOpen
  \bibfield  {author} {\bibinfo {author} {\bibfnamefont {J.-C.}\ \bibnamefont
  {Chervin}}, \bibinfo {author} {\bibfnamefont {B.}~\bibnamefont {Canny}}, \
  and\ \bibinfo {author} {\bibfnamefont {M.}~\bibnamefont {Mancinelli}},\
  }\href {\doibase 10.1080/08957950108202589} {\bibfield  {journal} {\bibinfo
  {journal} {High Pressure Res.}\ }\textbf {\bibinfo {volume} {21}},\ \bibinfo
  {pages} {305} (\bibinfo {year} {2001})}\BibitemShut {NoStop}%
\bibitem [{\citenamefont {Lindsay}\ \emph {et~al.}(1981)\citenamefont
  {Lindsay}, \citenamefont {Anderson},\ and\ \citenamefont
  {Sandercock}}]{LIND1981}%
  \BibitemOpen
  \bibfield  {author} {\bibinfo {author} {\bibfnamefont {S.~M.}\ \bibnamefont
  {Lindsay}}, \bibinfo {author} {\bibfnamefont {M.~W.}\ \bibnamefont
  {Anderson}}, \ and\ \bibinfo {author} {\bibfnamefont {J.~R.}\ \bibnamefont
  {Sandercock}},\ }\href {\doibase 10.1063/1.1136479} {\bibfield  {journal}
  {\bibinfo  {journal} {Rev. Sci. Instrum.}\ }\textbf {\bibinfo {volume}
  {52}},\ \bibinfo {pages} {1478} (\bibinfo {year} {1981})}\BibitemShut
  {NoStop}%
\bibitem [{\citenamefont {Vacher}\ \emph {et~al.}(2006)\citenamefont {Vacher},
  \citenamefont {Ayrinhac}, \citenamefont {Foret}, \citenamefont {Ruffl\'e},\
  and\ \citenamefont {Courtens}}]{VACH2006}%
  \BibitemOpen
  \bibfield  {author} {\bibinfo {author} {\bibfnamefont {R.}~\bibnamefont
  {Vacher}}, \bibinfo {author} {\bibfnamefont {S.}~\bibnamefont {Ayrinhac}},
  \bibinfo {author} {\bibfnamefont {M.}~\bibnamefont {Foret}}, \bibinfo
  {author} {\bibfnamefont {B.}~\bibnamefont {Ruffl\'e}}, \ and\ \bibinfo
  {author} {\bibfnamefont {E.}~\bibnamefont {Courtens}},\ }\href {\doibase
  10.1103/PhysRevB.74.012203} {\bibfield  {journal} {\bibinfo  {journal} {Phys.
  Rev. B}\ }\textbf {\bibinfo {volume} {74}},\ \bibinfo {pages} {012203}
  (\bibinfo {year} {2006})}\BibitemShut {NoStop}%
\bibitem [{\citenamefont {Anderson}\ and\ \citenamefont
  {B\"ommel}()}]{ANDE1955}%
  \BibitemOpen
  \bibfield  {author} {\bibinfo {author} {\bibfnamefont {O.~L.}\ \bibnamefont
  {Anderson}}\ and\ \bibinfo {author} {\bibfnamefont {H.~E.}\ \bibnamefont
  {B\"ommel}},\ }\href {\doibase 10.1111/j.1151-2916.1955.tb14914.x} {\bibfield
   {journal} {\bibinfo  {journal} {J. Am. Ceram. Soc.}\ }\textbf {\bibinfo
  {volume} {38}},\ \bibinfo {pages} {125}}\BibitemShut {NoStop}%
\bibitem [{\citenamefont {J\"ackle}\ \emph {et~al.}(1976)\citenamefont
  {J\"ackle}, \citenamefont {Pich\'e}, \citenamefont {Arnold},\ and\
  \citenamefont {Hunklinger}}]{JACK1976}%
  \BibitemOpen
  \bibfield  {author} {\bibinfo {author} {\bibfnamefont {J.}~\bibnamefont
  {J\"ackle}}, \bibinfo {author} {\bibfnamefont {L.}~\bibnamefont {Pich\'e}},
  \bibinfo {author} {\bibfnamefont {W.}~\bibnamefont {Arnold}}, \ and\ \bibinfo
  {author} {\bibfnamefont {S.}~\bibnamefont {Hunklinger}},\ }\href {\doibase
  10.1016/0022-3093(76)90119-8} {\bibfield  {journal} {\bibinfo  {journal} {J.
  Non-Cryst. Solids}\ }\textbf {\bibinfo {volume} {20}},\ \bibinfo {pages}
  {365} (\bibinfo {year} {1976})}\BibitemShut {NoStop}%
\bibitem [{\citenamefont {Maris}(1971)}]{MARI1971}%
  \BibitemOpen
  \bibfield  {author} {\bibinfo {author} {\bibfnamefont {H.~J.}\ \bibnamefont
  {Maris}},\ }in\ \href@noop {} {\emph {\bibinfo {booktitle} {Physical
  Acoustics}}},\ Vol.\ \bibinfo {volume} {VIII},\ \bibinfo {editor} {edited by\
  \bibinfo {editor} {\bibfnamefont {W.}~\bibnamefont {Mason}}\ and\ \bibinfo
  {editor} {\bibfnamefont {R.}~\bibnamefont {Thurston}}}\ (\bibinfo
  {publisher} {Academic Press},\ \bibinfo {address} {New York, NY},\ \bibinfo
  {year} {1971})\ p.\ \bibinfo {pages} {279}\BibitemShut {NoStop}%
\bibitem [{\citenamefont {Vacher}\ \emph {et~al.}(2005)\citenamefont {Vacher},
  \citenamefont {Courtens},\ and\ \citenamefont {Foret}}]{VACH2005}%
  \BibitemOpen
  \bibfield  {author} {\bibinfo {author} {\bibfnamefont {R.}~\bibnamefont
  {Vacher}}, \bibinfo {author} {\bibfnamefont {E.}~\bibnamefont {Courtens}}, \
  and\ \bibinfo {author} {\bibfnamefont {M.}~\bibnamefont {Foret}},\ }\href
  {\doibase 10.1103/PhysRevB.72.214205} {\bibfield  {journal} {\bibinfo
  {journal} {Phys. Rev. B}\ }\textbf {\bibinfo {volume} {72}},\ \bibinfo
  {pages} {214205} (\bibinfo {year} {2005})}\BibitemShut {NoStop}%
\bibitem [{\citenamefont {Ayrinhac}\ \emph {et~al.}(2011)\citenamefont
  {Ayrinhac}, \citenamefont {Ruffl\'e}, \citenamefont {Foret}, \citenamefont
  {Tran}, \citenamefont {Cl\'ement}, \citenamefont {Vialla}, \citenamefont
  {Vacher}, \citenamefont {Chervin}, \citenamefont {Munsch},\ and\
  \citenamefont {Polian}}]{AYRI2011a}%
  \BibitemOpen
  \bibfield  {author} {\bibinfo {author} {\bibfnamefont {S.}~\bibnamefont
  {Ayrinhac}}, \bibinfo {author} {\bibfnamefont {B.}~\bibnamefont {Ruffl\'e}},
  \bibinfo {author} {\bibfnamefont {M.}~\bibnamefont {Foret}}, \bibinfo
  {author} {\bibfnamefont {H.}~\bibnamefont {Tran}}, \bibinfo {author}
  {\bibfnamefont {S.}~\bibnamefont {Cl\'ement}}, \bibinfo {author}
  {\bibfnamefont {R.}~\bibnamefont {Vialla}}, \bibinfo {author} {\bibfnamefont
  {R.}~\bibnamefont {Vacher}}, \bibinfo {author} {\bibfnamefont {J.~C.}\
  \bibnamefont {Chervin}}, \bibinfo {author} {\bibfnamefont {P.}~\bibnamefont
  {Munsch}}, \ and\ \bibinfo {author} {\bibfnamefont {A.}~\bibnamefont
  {Polian}},\ }\href {\doibase 10.1103/PhysRevB.84.024201} {\bibfield
  {journal} {\bibinfo  {journal} {Phys. Rev. B}\ }\textbf {\bibinfo {volume}
  {84}},\ \bibinfo {pages} {024201} (\bibinfo {year} {2011})}\BibitemShut
  {NoStop}%
\bibitem [{\citenamefont {Ruffl\'e}\ \emph {et~al.}(2010)\citenamefont
  {Ruffl\'e}, \citenamefont {Ayrinhac}, \citenamefont {Courtens}, \citenamefont
  {Vacher}, \citenamefont {Foret}, \citenamefont {Wischnewski},\ and\
  \citenamefont {Buchenau}}]{RUFF2010}%
  \BibitemOpen
  \bibfield  {author} {\bibinfo {author} {\bibfnamefont {B.}~\bibnamefont
  {Ruffl\'e}}, \bibinfo {author} {\bibfnamefont {S.}~\bibnamefont {Ayrinhac}},
  \bibinfo {author} {\bibfnamefont {E.}~\bibnamefont {Courtens}}, \bibinfo
  {author} {\bibfnamefont {R.}~\bibnamefont {Vacher}}, \bibinfo {author}
  {\bibfnamefont {M.}~\bibnamefont {Foret}}, \bibinfo {author} {\bibfnamefont
  {A.}~\bibnamefont {Wischnewski}}, \ and\ \bibinfo {author} {\bibfnamefont
  {U.}~\bibnamefont {Buchenau}},\ }\href {\doibase
  10.1103/PhysRevLett.104.067402} {\bibfield  {journal} {\bibinfo  {journal}
  {Phys. Rev. Lett.}\ }\textbf {\bibinfo {volume} {104}},\ \bibinfo {pages}
  {067402} (\bibinfo {year} {2010})}\BibitemShut {NoStop}%
\bibitem [{\citenamefont {Petitgirard}\ \emph {et~al.}(2017)\citenamefont
  {Petitgirard}, \citenamefont {Malfait}, \citenamefont {Journaux},
  \citenamefont {Collings}, \citenamefont {Jennings}, \citenamefont
  {Blanchard}, \citenamefont {Kantor}, \citenamefont {Kurnosov}, \citenamefont
  {Cotte}, \citenamefont {Dane}, \citenamefont {Burghammer},\ and\
  \citenamefont {Rubie}}]{PETI2017}%
  \BibitemOpen
  \bibfield  {author} {\bibinfo {author} {\bibfnamefont {S.}~\bibnamefont
  {Petitgirard}}, \bibinfo {author} {\bibfnamefont {W.~J.}\ \bibnamefont
  {Malfait}}, \bibinfo {author} {\bibfnamefont {B.}~\bibnamefont {Journaux}},
  \bibinfo {author} {\bibfnamefont {I.~E.}\ \bibnamefont {Collings}}, \bibinfo
  {author} {\bibfnamefont {E.~S.}\ \bibnamefont {Jennings}}, \bibinfo {author}
  {\bibfnamefont {I.}~\bibnamefont {Blanchard}}, \bibinfo {author}
  {\bibfnamefont {I.}~\bibnamefont {Kantor}}, \bibinfo {author} {\bibfnamefont
  {A.}~\bibnamefont {Kurnosov}}, \bibinfo {author} {\bibfnamefont
  {M.}~\bibnamefont {Cotte}}, \bibinfo {author} {\bibfnamefont
  {T.}~\bibnamefont {Dane}}, \bibinfo {author} {\bibfnamefont {M.}~\bibnamefont
  {Burghammer}}, \ and\ \bibinfo {author} {\bibfnamefont {D.~C.}\ \bibnamefont
  {Rubie}},\ }\href {https://link.aps.org/doi/10.1103/PhysRevLett.119.215701}
  {\bibfield  {journal} {\bibinfo  {journal} {Phys. Rev. Lett.}\ }\textbf
  {\bibinfo {volume} {119}},\ \bibinfo {pages} {215701} (\bibinfo {year}
  {2017})}\BibitemShut {NoStop}%
\bibitem [{\citenamefont {Sato}\ \emph {et~al.}(2018)\citenamefont {Sato},
  \citenamefont {Funamori}, \citenamefont {Wakabayashi}, \citenamefont
  {Nishida},\ and\ \citenamefont {Kikegawa}}]{SATO2018}%
  \BibitemOpen
  \bibfield  {author} {\bibinfo {author} {\bibfnamefont {T.}~\bibnamefont
  {Sato}}, \bibinfo {author} {\bibfnamefont {N.}~\bibnamefont {Funamori}},
  \bibinfo {author} {\bibfnamefont {D.}~\bibnamefont {Wakabayashi}}, \bibinfo
  {author} {\bibfnamefont {K.}~\bibnamefont {Nishida}}, \ and\ \bibinfo
  {author} {\bibfnamefont {T.}~\bibnamefont {Kikegawa}},\ }\href
  {https://link.aps.org/doi/10.1103/PhysRevB.98.144111} {\bibfield  {journal}
  {\bibinfo  {journal} {Phys. Rev. B}\ }\textbf {\bibinfo {volume} {98}},\
  \bibinfo {pages} {144111} (\bibinfo {year} {2018})}\BibitemShut {NoStop}%
\bibitem [{\citenamefont {Ryuo}\ \emph {et~al.}(2017)\citenamefont {Ryuo},
  \citenamefont {Wakabayashi}, \citenamefont {Koura},\ and\ \citenamefont
  {Shimojo}}]{RYUO2017}%
  \BibitemOpen
  \bibfield  {author} {\bibinfo {author} {\bibfnamefont {E.}~\bibnamefont
  {Ryuo}}, \bibinfo {author} {\bibfnamefont {D.}~\bibnamefont {Wakabayashi}},
  \bibinfo {author} {\bibfnamefont {A.}~\bibnamefont {Koura}}, \ and\ \bibinfo
  {author} {\bibfnamefont {F.}~\bibnamefont {Shimojo}},\ }\href
  {https://link.aps.org/doi/10.1103/PhysRevB.96.054206} {\bibfield  {journal}
  {\bibinfo  {journal} {Phys. Rev. B}\ }\textbf {\bibinfo {volume} {96}},\
  \bibinfo {pages} {054206} (\bibinfo {year} {2017})}\BibitemShut {NoStop}%
\bibitem [{\citenamefont {Nicholas}\ \emph {et~al.}(2004)\citenamefont
  {Nicholas}, \citenamefont {Sinogeikin}, \citenamefont {Kieffer},\ and\
  \citenamefont {Bass}}]{NICH2004}%
  \BibitemOpen
  \bibfield  {author} {\bibinfo {author} {\bibfnamefont {J.}~\bibnamefont
  {Nicholas}}, \bibinfo {author} {\bibfnamefont {S.}~\bibnamefont
  {Sinogeikin}}, \bibinfo {author} {\bibfnamefont {J.}~\bibnamefont {Kieffer}},
  \ and\ \bibinfo {author} {\bibfnamefont {J.}~\bibnamefont {Bass}},\ }\href
  {https://link.aps.org/doi/10.1103/PhysRevLett.92.215701} {\bibfield
  {journal} {\bibinfo  {journal} {Phys. Rev. Lett.}\ }\textbf {\bibinfo
  {volume} {92}},\ \bibinfo {pages} {215701} (\bibinfo {year}
  {2004})}\BibitemShut {NoStop}%
\bibitem [{\citenamefont {Huang}\ \emph {et~al.}(2008)\citenamefont {Huang},
  \citenamefont {Nicholas}, \citenamefont {Kieffer},\ and\ \citenamefont
  {Bass}}]{HUAN2008}%
  \BibitemOpen
  \bibfield  {author} {\bibinfo {author} {\bibfnamefont {L.}~\bibnamefont
  {Huang}}, \bibinfo {author} {\bibfnamefont {J.}~\bibnamefont {Nicholas}},
  \bibinfo {author} {\bibfnamefont {J.}~\bibnamefont {Kieffer}}, \ and\
  \bibinfo {author} {\bibfnamefont {J.}~\bibnamefont {Bass}},\ }\href {\doibase
  10.1088/0953-8984/20/7/075107} {\bibfield  {journal} {\bibinfo  {journal} {J.
  Phys.-Cond. Matter}\ }\textbf {\bibinfo {volume} {20}},\ \bibinfo {pages}
  {075107} (\bibinfo {year} {2008})}\BibitemShut {NoStop}%
\bibitem [{\citenamefont {Zeidler}\ \emph
  {et~al.}(2014{\natexlab{b}})\citenamefont {Zeidler}, \citenamefont {Wezka},
  \citenamefont {Whittaker}, \citenamefont {Salmon}, \citenamefont {Baroni},
  \citenamefont {Klotz}, \citenamefont {Fischer}, \citenamefont {Wilding},
  \citenamefont {Bull}, \citenamefont {Tucker}, \citenamefont {Salanne},
  \citenamefont {Ferlat},\ and\ \citenamefont {Micoulaut}}]{ZEID2014a}%
  \BibitemOpen
  \bibfield  {author} {\bibinfo {author} {\bibfnamefont {A.}~\bibnamefont
  {Zeidler}}, \bibinfo {author} {\bibfnamefont {K.}~\bibnamefont {Wezka}},
  \bibinfo {author} {\bibfnamefont {D.~A.~J.}\ \bibnamefont {Whittaker}},
  \bibinfo {author} {\bibfnamefont {P.~S.}\ \bibnamefont {Salmon}}, \bibinfo
  {author} {\bibfnamefont {A.}~\bibnamefont {Baroni}}, \bibinfo {author}
  {\bibfnamefont {S.}~\bibnamefont {Klotz}}, \bibinfo {author} {\bibfnamefont
  {H.~E.}\ \bibnamefont {Fischer}}, \bibinfo {author} {\bibfnamefont {M.~C.}\
  \bibnamefont {Wilding}}, \bibinfo {author} {\bibfnamefont {C.~L.}\
  \bibnamefont {Bull}}, \bibinfo {author} {\bibfnamefont {M.~G.}\ \bibnamefont
  {Tucker}}, \bibinfo {author} {\bibfnamefont {M.}~\bibnamefont {Salanne}},
  \bibinfo {author} {\bibfnamefont {G.}~\bibnamefont {Ferlat}}, \ and\ \bibinfo
  {author} {\bibfnamefont {M.}~\bibnamefont {Micoulaut}},\ }\href
  {https://link.aps.org/doi/10.1103/PhysRevB.90.024206} {\bibfield  {journal}
  {\bibinfo  {journal} {Phys. Rev. B}\ }\textbf {\bibinfo {volume} {90}},\
  \bibinfo {pages} {024206} (\bibinfo {year} {2014}{\natexlab{b}})}\BibitemShut
  {NoStop}%
\bibitem [{\citenamefont {Brazhkin}\ \emph {et~al.}(2009)\citenamefont
  {Brazhkin}, \citenamefont {Tsiok},\ and\ \citenamefont
  {Katayama}}]{BRAZ2009a}%
  \BibitemOpen
  \bibfield  {author} {\bibinfo {author} {\bibfnamefont {V.~V.}\ \bibnamefont
  {Brazhkin}}, \bibinfo {author} {\bibfnamefont {O.~B.}\ \bibnamefont {Tsiok}},
  \ and\ \bibinfo {author} {\bibfnamefont {Y.}~\bibnamefont {Katayama}},\
  }\href {\doibase 10.1134/S0021364009050063} {\bibfield  {journal} {\bibinfo
  {journal} {JETP Letters}\ }\textbf {\bibinfo {volume} {89}},\ \bibinfo
  {pages} {244} (\bibinfo {year} {2009})}\BibitemShut {NoStop}%
\bibitem [{\citenamefont {Rouxel}\ \emph {et~al.}(2008)\citenamefont {Rouxel},
  \citenamefont {Ji}, \citenamefont {Hammouda},\ and\ \citenamefont
  {Mor\'eac}}]{ROUX2008}%
  \BibitemOpen
  \bibfield  {author} {\bibinfo {author} {\bibfnamefont {T.}~\bibnamefont
  {Rouxel}}, \bibinfo {author} {\bibfnamefont {H.}~\bibnamefont {Ji}}, \bibinfo
  {author} {\bibfnamefont {T.}~\bibnamefont {Hammouda}}, \ and\ \bibinfo
  {author} {\bibfnamefont {A.}~\bibnamefont {Mor\'eac}},\ }\href {\doibase
  10.1103/PhysRevLett.100.225501} {\bibfield  {journal} {\bibinfo  {journal}
  {Phys. Rev. Lett.}\ }\textbf {\bibinfo {volume} {100}},\ \bibinfo {pages}
  {225501} (\bibinfo {year} {2008})}\BibitemShut {NoStop}%
\bibitem [{\citenamefont {Sundararaman}\ \emph {et~al.}(2018)\citenamefont
  {Sundararaman}, \citenamefont {Huang}, \citenamefont {Ispas},\ and\
  \citenamefont {Kob}}]{SUND2018}%
  \BibitemOpen
  \bibfield  {author} {\bibinfo {author} {\bibfnamefont {S.}~\bibnamefont
  {Sundararaman}}, \bibinfo {author} {\bibfnamefont {L.}~\bibnamefont {Huang}},
  \bibinfo {author} {\bibfnamefont {S.}~\bibnamefont {Ispas}}, \ and\ \bibinfo
  {author} {\bibfnamefont {W.}~\bibnamefont {Kob}},\ }\href
  {https://doi.org/10.1063/1.5023707} {\bibfield  {journal} {\bibinfo
  {journal} {J. Chem. Phys.}\ }\textbf {\bibinfo {volume} {148}},\ \bibinfo
  {pages} {194504} (\bibinfo {year} {2018})}\BibitemShut {NoStop}%
\bibitem [{\citenamefont {Liang}\ \emph {et~al.}(2007)\citenamefont {Liang},
  \citenamefont {Miranda},\ and\ \citenamefont {Scandolo}}]{LIAN2007}%
  \BibitemOpen
  \bibfield  {author} {\bibinfo {author} {\bibfnamefont {Y.}~\bibnamefont
  {Liang}}, \bibinfo {author} {\bibfnamefont {C.~R.}\ \bibnamefont {Miranda}},
  \ and\ \bibinfo {author} {\bibfnamefont {S.}~\bibnamefont {Scandolo}},\
  }\href@noop {} {\bibfield  {journal} {\bibinfo  {journal} {Phys. Rev. B}\
  }\textbf {\bibinfo {volume} {75}},\ \bibinfo {pages} {024205} (\bibinfo
  {year} {2007})}\BibitemShut {NoStop}%
\bibitem [{\citenamefont {Mantisi}\ \emph {et~al.}(2012)\citenamefont
  {Mantisi}, \citenamefont {Tanguy}, \citenamefont {Kermouche},\ and\
  \citenamefont {Barthel}}]{MANT2012}%
  \BibitemOpen
  \bibfield  {author} {\bibinfo {author} {\bibfnamefont {B.}~\bibnamefont
  {Mantisi}}, \bibinfo {author} {\bibfnamefont {A.}~\bibnamefont {Tanguy}},
  \bibinfo {author} {\bibfnamefont {G.}~\bibnamefont {Kermouche}}, \ and\
  \bibinfo {author} {\bibfnamefont {E.}~\bibnamefont {Barthel}},\ }\href
  {\doibase 10.1140/epjb/e2012-30317-6} {\bibfield  {journal} {\bibinfo
  {journal} {Eur. Phys. J. B}\ }\textbf {\bibinfo {volume} {85}},\ \bibinfo
  {pages} {304} (\bibinfo {year} {2012})}\BibitemShut {NoStop}%
\bibitem [{\citenamefont {Mackenzie}(1963)}]{MACK1963}%
  \BibitemOpen
  \bibfield  {author} {\bibinfo {author} {\bibfnamefont {J.}~\bibnamefont
  {Mackenzie}},\ }\href@noop {} {\bibfield  {journal} {\bibinfo  {journal} {J.
  Am. Ceram. Soc.}\ }\textbf {\bibinfo {volume} {46}},\ \bibinfo {pages} {461}
  (\bibinfo {year} {1963})}\BibitemShut {NoStop}%
\bibitem [{\citenamefont {Rat}\ \emph {et~al.}(1999)\citenamefont {Rat},
  \citenamefont {Foret}, \citenamefont {Courtens}, \citenamefont {Vacher},\
  and\ \citenamefont {Arai}}]{RAT1999}%
  \BibitemOpen
  \bibfield  {author} {\bibinfo {author} {\bibfnamefont {E.}~\bibnamefont
  {Rat}}, \bibinfo {author} {\bibfnamefont {M.}~\bibnamefont {Foret}}, \bibinfo
  {author} {\bibfnamefont {E.}~\bibnamefont {Courtens}}, \bibinfo {author}
  {\bibfnamefont {R.}~\bibnamefont {Vacher}}, \ and\ \bibinfo {author}
  {\bibfnamefont {M.}~\bibnamefont {Arai}},\ }\href@noop {} {\bibfield
  {journal} {\bibinfo  {journal} {Phys. Rev. Lett.}\ }\textbf {\bibinfo
  {volume} {83}},\ \bibinfo {pages} {1355} (\bibinfo {year}
  {1999})}\BibitemShut {NoStop}%
\bibitem [{\citenamefont {Guerette}\ \emph {et~al.}(2015)\citenamefont
  {Guerette}, \citenamefont {Ackerson}, \citenamefont {Thomas}, \citenamefont
  {Yuan}, \citenamefont {Watson}, \citenamefont {Walker},\ and\ \citenamefont
  {Huang}}]{GUER2015}%
  \BibitemOpen
  \bibfield  {author} {\bibinfo {author} {\bibfnamefont {M.}~\bibnamefont
  {Guerette}}, \bibinfo {author} {\bibfnamefont {M.~R.}\ \bibnamefont
  {Ackerson}}, \bibinfo {author} {\bibfnamefont {J.}~\bibnamefont {Thomas}},
  \bibinfo {author} {\bibfnamefont {F.}~\bibnamefont {Yuan}}, \bibinfo {author}
  {\bibfnamefont {E.~B.}\ \bibnamefont {Watson}}, \bibinfo {author}
  {\bibfnamefont {D.}~\bibnamefont {Walker}}, \ and\ \bibinfo {author}
  {\bibfnamefont {L.}~\bibnamefont {Huang}},\ }\href {\doibase
  {10.1038/srep15343}} {\bibfield  {journal} {\bibinfo  {journal} {{Scientific
  Reports}}\ }\textbf {\bibinfo {volume} {{5}}} (\bibinfo {year} {{2015}}),\
  {10.1038/srep15343}}\BibitemShut {NoStop}%
\bibitem [{\citenamefont {Takada}(2018)}]{TAKA2018}%
  \BibitemOpen
  \bibfield  {author} {\bibinfo {author} {\bibfnamefont {A.}~\bibnamefont
  {Takada}},\ }\href {\doibase
  https://doi.org/10.1016/j.jnoncrysol.2018.07.037} {\bibfield  {journal}
  {\bibinfo  {journal} {J. Non-Cryst. Solids}\ }\textbf {\bibinfo {volume}
  {499}},\ \bibinfo {pages} {309 } (\bibinfo {year} {2018})}\BibitemShut
  {NoStop}%
\bibitem [{\citenamefont {Brazhkin}\ \emph {et~al.}(2011)\citenamefont
  {Brazhkin}, \citenamefont {Lyapin},\ and\ \citenamefont
  {Trachenko}}]{BRAZ2011}%
  \BibitemOpen
  \bibfield  {author} {\bibinfo {author} {\bibfnamefont {V.~V.}\ \bibnamefont
  {Brazhkin}}, \bibinfo {author} {\bibfnamefont {A.~G.}\ \bibnamefont
  {Lyapin}}, \ and\ \bibinfo {author} {\bibfnamefont {K.}~\bibnamefont
  {Trachenko}},\ }\href {\doibase 10.1103/PhysRevB.83.132103} {\bibfield
  {journal} {\bibinfo  {journal} {Phys. Rev. B}\ }\textbf {\bibinfo {volume}
  {83}},\ \bibinfo {pages} {132103} (\bibinfo {year} {2011})}\BibitemShut
  {NoStop}%
\bibitem [{\citenamefont {Kono}\ \emph {et~al.}(2016)\citenamefont {Kono},
  \citenamefont {Kenney-Benson}, \citenamefont {Ikuta}, \citenamefont
  {Shibazaki}, \citenamefont {Wang},\ and\ \citenamefont {Shen}}]{KONO2016}%
  \BibitemOpen
  \bibfield  {author} {\bibinfo {author} {\bibfnamefont {Y.}~\bibnamefont
  {Kono}}, \bibinfo {author} {\bibfnamefont {C.}~\bibnamefont {Kenney-Benson}},
  \bibinfo {author} {\bibfnamefont {D.}~\bibnamefont {Ikuta}}, \bibinfo
  {author} {\bibfnamefont {Y.}~\bibnamefont {Shibazaki}}, \bibinfo {author}
  {\bibfnamefont {Y.}~\bibnamefont {Wang}}, \ and\ \bibinfo {author}
  {\bibfnamefont {G.}~\bibnamefont {Shen}},\ }\href {\doibase
  10.1073/pnas.1524304113} {\bibfield  {journal} {\bibinfo  {journal} {Proc.
  Natl. Acad. Sci.}\ }\textbf {\bibinfo {volume} {113}},\ \bibinfo {pages}
  {3436} (\bibinfo {year} {2016})}\BibitemShut {NoStop}%
\end{thebibliography}
%

\end{document}